\documentclass[a4paper,11pt]{article}
\pdfoutput=1 % if your are submitting a pdflatex (i.e. if you have
\usepackage{jinstpub}
\usepackage{enumitem}

\title{Environment-friendly gas mixtures for Resistive Plate Chambers: an experimental and simulation study}
\author[a,1]{A. Bianchi,\note{Corresponding author.}}
\author[a]{S. Delsanto,}
\author[b]{P. Dupieux,}
\author[a]{A. Ferretti,}
\author[a]{M. Gagliardi,}
\author[b]{B. Joly,}
\author[b]{S. P. Manen,}
\author[c]{M. Marchisone,}
\author[a]{L. Micheletti,}
\author[a]{L. Quaglia,}
\author[a]{A. Rosano,}
\author[a]{L. Terlizzi}
\author[a]{and E. Vercellin}

\affiliation[a]{Universit\`a degli Studi di Torino and INFN, Sezione di Torino, Via Pietro Giuria 1, 10125, Torino, Italy}
\affiliation[b]{Clermont Universit\'e, Universit\'e Blaise Pascal, CNRS/IN2P3, Laboratoire de Physique Corpusculaire, BP 10448, F-63000 Clermont-Ferrand, France}
\affiliation[c]{Institut de Physique Nucl\'eaire de Lyon, Universit\'e Claude Bernard, 4 rue Enrico Fermi, 69622, Villeurbanne, France}

\emailAdd{antonio.bianchi@unito.it}

\abstract{
Resistive Plate Chambers (RPC) have shown stable operation at the Large Hadron Collider and satisfactory efficiency for the entire Run 1 (2010-2013) and Run 2 (2015-2018) with C$_{2}$H$_{2}$F$_{4}$-based gas mixtures and the addition of SF$_{6}$ and \textit{i}-C$_{4}$H$_{10}$. Since its global warming potential (GWP) is high, C$_{2}$H$_{2}$F$_{4}$ is phasing out of production due to recent European Union regulations and as a result its cost is progressively increasing. Therefore, finding a new RPC gas mixture with a low GWP has become extremely important.

This contribution describes the simulation of the RPC efficiency with tetrafluoropropene C$_{3}$H$_{2}$F$_{4}$ (HFO1234ze), a hydrofluoroolefin with very low GWP. Simulation results are systematically compared with measurements of RPC efficiency in C$_{3}$H$_{2}$F$_{4}$-based gas mixtures with the addition of different combinations of Ar, He, CO$_{2}$, O$_{2}$ and \textit{i}-C$_{4}$H$_{10}$ in various concentrations. This simulation allows the study of the interplay between C$_{3}$H$_{2}$F$_{4}$ and the other gas components in the mixture as well as may allow the identification of the most promising environment-friendly gas mixtures with C$_{3}$H$_{2}$F$_{4}$ for RPCs.
}


\proceeding{XV Workshop on Resistive Plate Chambers and Related Detectors (RPC2020)\\
  when 10-14 February 2020\\
  where Roma, Italy}

\begin{document}
\maketitle
\flushbottom

\section{Introduction}
Resistive Plate Chambers (RPC) have shown stable operation at the Large Hadron Collider (LHC) and satisfactory efficiency for the entire Run 1 (2010-2013) and Run 2 (2015-2018) with C$_{2}$H$_{2}$F$_{4}$-based gas mixtures and the addition of \textit{i}-C$_{4}$H$_{10}$ and SF$_{6}$. The RPC gas mixture at the ATLAS and CMS experiments contains 4.5\% \textit{i}-C$_{4}$H$_{10}$ and 0.3\% SF$_{6}$ while the concentrations of \textit{i}-C$_{4}$H$_{10}$ and SF$_{6}$ in the ALICE gas mixture are 10.0\% and 0.3\%, respectively. 

The heat trapped by a greenhouse gas in the atmosphere is assessed by its global warming potential (GWP). The GWP of CO$_{2}$ is equal to 1 by definition while this potential is even more than three orders of magnitude for C$_{2}$H$_{2}$F$_{4}$ (GWP = \textasciitilde1430 \cite{europeanparliament_ok}). Since the concentration of C$_{2}$H$_{2}$F$_{4}$ in RPC gas mixtures is high (between 89\% and 95\%), more than 95\% of their total GWP is due to the presence of C$_{2}$H$_{2}$F$_{4}$.

Recent regulations from the European Union (EU) impose a gradual limitation of the production of fluorinated greenhouse gases (such as C$_{2}$H$_{2}$F$_{4}$) \cite{europeanparliament_ok}. As a result, their cost is progressively increasing. In parallel, CERN has elaborated a number of strategies to reduce as much as possible the emissions of greenhouse gases or, at least, optimize their use in the LHC experiments \cite{capeans2017strategies_ok}. For these reasons, finding a new gas mixture with a low GWP for RPCs has become extremely important.

The usage of tetrafluoropropene C$_{3}$H$_{2}$F$_{4}$ (HFO1234ze) for RPC gas mixtures has widely been explored in the recent years as this gas may represent an environment-friendly alternative to C$_{2}$H$_{2}$F$_{4}$ \cite{cardarelli_ok, Abbrescia_2016_He_ok, guida2016characterization_ok, liberti2016further_ok, bianchi2019r_ok, bianchi2019characterization_ok, bianchi2020}. Indeed, the GWP of C$_{3}$H$_{2}$F$_{4}$ is less than 1 according to the last Intergovernmental Panel on Climate Change (IPCC) Assessment Report \cite{IPCCclimate_ok}. As far as we know, all findings on eco-friendly gas mixture for RPCs have exclusively been obtained by an experimental approach so far \cite{cardarelli_ok, Abbrescia_2016_He_ok, guida2016characterization_ok, liberti2016further_ok, bianchi2019r_ok, bianchi2019characterization_ok, bianchi2020}. Indeed, the lack of knowledge on fundamental parameters of C$_{3}$H$_{2}$F$_{4}$, e.g. its electron collision cross sections, makes the implementation of this gas in simulations rather difficult \cite{bianchi2019characterization_ok}. 

One of the main issue in the identification of a suitable RPC gas mixture with a low environmental impact is the operating voltage of the detector that must be within a limited high voltage ($HV$) range \cite{bianchi2019characterization_ok}. This implies the need to understand how the RPC performance and, in particular, its efficiency curve as a function of the $HV$ change due to the interplay between C$_{3}$H$_{2}$F$_{4}$ and the other gas components in the mixture.

In this work we report a simulation study of the RPC efficiency in several C$_{3}$H$_{2}$F$_{4}$-based gas mixtures with the addition of various gases in different concentrations. Simulation results are systematically compared with our measurements \cite{bianchi2019characterization_ok, bianchi2020} in C$_{3}$H$_{2}$F$_{4}$-based gas mixtures in order to validate the simulation and obtain a reliable tool to identify the most promising gas mixtures with a low GWP for RPCs.

The paper is organized as follows. In section~\ref{sec:simulazioneREFF} we describe the REFF (RPC EFFiciency) simulation and its validation. A systematic comparison between simulation and measurement of the RPC efficiency in C$_{3}$H$_{2}$F$_{4}$-based gas mixtures is presented in section~\ref{sec:confrontodatiesimulazione}. Finally, conclusions are drawn in section~\ref{sec:conclusioni}.

\section{REFF simulation and its validation}\label{sec:simulazioneREFF}
The REFF simulation implements a simplified method to evaluate the RPC efficiency as a function of the $HV$ in different gas mixtures. Indeed, the efficiency is assessed by simulating the avalanche size at the anode, originated by the primary electrons in the gas gap of the detector. These primary electrons are released by the passage of an incoming radiation due to the ionization of the gas mixture in the detector. %This simulation procedure is rather similar to the so-called \textit{1.5D-model}, developed by Lippmann and Riegler \cite{riegler_2002, riegler_2004, lippmann_2004}.

Input values of the REFF simulation are the volume composition of the gas mixture and its density, the type and energy of the ionizing particle, and the effective ionization Townsend coefficient for each \textit{HV} value of interest. The initial position of clusters in the gas gap and the number of primary electrons in each cluster are evaluated by HEED \cite{codeHEED}, which is a dedicated toolkit for this purpose. The number of primary electrons in C$_{3}$H$_{2}$F$_{4}$ is assumed to be \textasciitilde9/mm according to the Benussi at al.'s simulation \cite{lavoroBenussi}. Concerning the effective ionization Townsend coefficient as a function of the \textit{HV}, this parameter is evaluated by a dedicated Monte Carlo simulation \cite{bianchi_PhDthesis} starting from the scattering cross sections of electrons in the gas mixture of interest. For all gases, except for C$_{3}$H$_{2}$F$_{4}$, we use the electron collision cross sections provided by Biagi \cite{webLXCat} as they are extensively tested in the MAGBOLTZ program \cite{Magboltz_Biagi}. On the contrary, the set of electron collision cross sections for C$_{3}$H$_{2}$F$_{4}$ has been obtained by an iterative method of unfolding the electron swarm parameters in C$_{3}$H$_{2}$F$_{4}$ \cite{bianchi_PhDthesis}, measured by Chachereau et al. \cite{HFO_ETH}. As far as we know, other sets of electron collision cross sections for C$_{3}$H$_{2}$F$_{4}$ are not currently available in the literature. More details on the iterative method as well as on the electron collision cross sections of C$_{3}$H$_{2}$F$_{4}$, used in this work, can be found in previous works \cite{bianchi_PhDthesis, bianchi_preparation}.

In the REFF simulation the avalanche development is modeled as an exponential growth. In particular, the electron mean free path $X$ is determined by the inverse of the effective ionization Townsend coefficient ($\alpha$\textsubscript{\textit{eff}}), which is the sum of ionization ($\alpha$) and attachment ($\eta$) Townsend coefficients. If the attachment coefficient cannot be determined, $\eta$ is assumed null while $\alpha$ becomes equal to $\alpha$\textsubscript{\textit{eff}}. Subsequently, the distance $\Delta x$ is defined as a fraction of the electron mean free path. In this work, $\Delta x$ is arbitrarily chosen two orders of magnitude lower than $X$. 

The charge multiplication in the avalanche is simulated by progressively increasing the trajectory of each electron in the gas gap along the direction of the electric field until all electrons reach the anode. In particular, the position of electrons is incremented by $\Delta x$ if a random number, generated from a uniform distribution in the interval from 0 to 1, is higher than $\frac{\Delta x}{X}$. On the contrary, if the condition is not verified, either an ionization or an attachment occurs. The selection of the right process to simulate is done by comparing an additional random number \textit{s} generated by the same uniform distribution with the ratio $\frac{\eta}{(\alpha + \eta)}$. If the condition $s \leq \frac{\eta}{(\alpha + \eta)}$ is verified, the electron is attached and thus the simulation of its trajectory in the gas gap is stopped, otherwise a new electron is added to the avalanche at the same position of the incident electron and its trajectory will be simulated until it is attached or reaches the anode. For simplicity, only electrons are considered in the REFF simulation, whereas ions are not taken into account as well as the deformation of the electric field lines due to space charge effects along the gas gap.

In order to calculate the detector efficiency for the gas mixture of interest and at a given value of $HV$, a number of different avalanche sizes at the anode are calculated and then compared to a threshold value of electrons that roughly represents the electronic threshold of the front-end electronics. This is a simplified way to evaluate the detector efficiency; alternatively, the Ramo-Shockley teorem \cite{teorema_Shockley, teorema_Ramo, riegler_2002} should be considered for an accurate description of the signal formation in RPCs.

The REFF simulation is validated by comparing its results with measurements of RPC efficiency as a function of the \textit{HV} at 1000 mbar and 293 K in 90\% C$_{2}$H$_{2}$F$_{4}$ and 10\% \textit{i}-C$_{4}$H$_{10}$ and also in gas mixtures of 55\% C$_{3}$H$_{2}$F$_{4}$ and the remaining fraction of Ar, CO$_{2}$ or O$_{2}$.

A dedicated experimental set-up with one small-size (50 $\times$ 50 cm\textsuperscript{2}, 2 mm single-gap) bakelite RPC has been used to measure the efficiency with different gas mixtures. The features of the RPC under test are similar to those of the ALICE-muon RPCs \cite{alice1999alice_ok, arnaldi2000alice_ok, c_ok} while signals are discriminated by the amplified front-end electronics FEERIC \cite{e_ok} with a threshold of \textasciitilde130 fC. A detailed description of the experimental set-up is reported in a previous work \cite{bianchi2019characterization_ok} while all measurements, presented in this paper, are already published in \cite{bianchi2019characterization_ok, bianchi2020}.

\begin{figure}[h]
    \centering
    \includegraphics[width=0.85\textwidth]{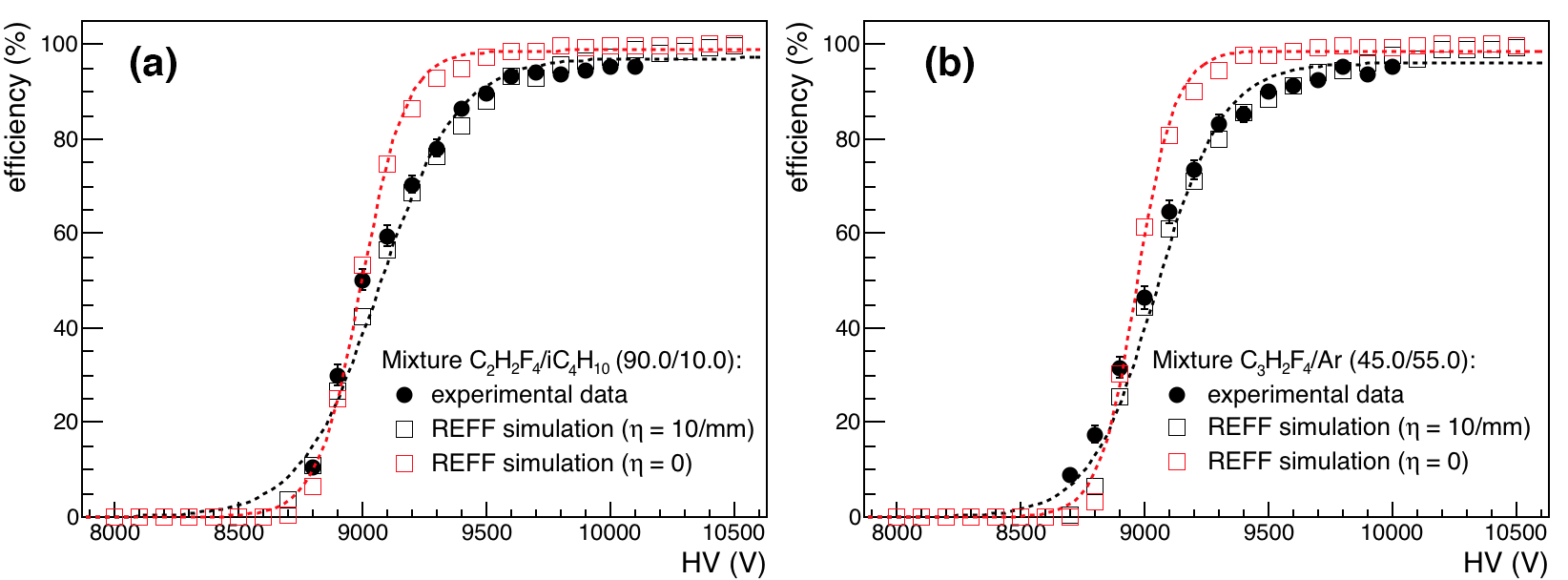}
    \caption{Efficiency curves obtained by the REFF simulation ($\eta$ = 0/mm and 10/mm) and the experimental data (a) in the gas mixture of 90\% C$_{2}$H$_{2}$F$_{4}$ and 10\% \textit{i}-C$_{4}$H$_{10}$ and (b) in the gas mixture of 55\% C$_{3}$H$_{2}$F$_{4}$ and 45\% Ar. Simulation results are interpolated by sigmoid functions in both plots. Some statistical error bars are hidden by markers.}
    \label{fig:plot0}
\end{figure}

Figure \ref{fig:plot0}a shows the simulation of the RPC efficiency curve in 90\% C$_{2}$H$_{2}$F$_{4}$ and 10\% \textit{i}-C$_{4}$H$_{10}$, evaluating the avalanche size of at the anode 5000 events for each $HV$ value ranging 8.5--10.5 kV. Simulations are carried out with $\alpha(HV)$ equal to the corresponding measurement of $\alpha$\textsubscript{\textit{eff}}($HV$), obtained by Colucci et al. \cite{exp_colucci} in a laser beam experiment, and $\eta(HV)$ equal to 0 mm$^{-1}$. The agreement between simulation results, calculated by assuming $\eta$ = 0 mm$^{-1}$, and experimental data is fairly good. In fact, \textit{HV} values at efficiency of 50\% differ in less than \textasciitilde50 V, but their difference at efficiency of 90\% is more than \textasciitilde300 V. It is important to highlight that the RPC efficiency does not only depend on $\alpha$\textsubscript{\textit{eff}} but it is affected by the value of both $\alpha$ and $\eta$, as already observed by Riegler et al. \cite{riegler_2002}. Since $\alpha$ and $\eta$ are not experimentally measured, the value of $\eta$ has been tuned from 0 to 20 mm$^{-1}$ in order to achieve a satisfactory agreement between simulation results and measurements. With $\eta(HV)$ equal to 10 mm$^{-1}$, differences between the simulation and the measurement are lower than 10\% for every \textit{HV} values, as shown in figure \ref{fig:plot0}a. As mentioned above, the RPC efficiency is measured by FEERIC with a discrimination threshold of \textasciitilde130 fC. This value corresponds to an induced signal of about 10$^{6}$ electrons, assuming the bakelite permittivity of 10 and the electrode thickness of 2 mm as in the case of the RPC under test \cite{riegler_2002, bianchi_PhDthesis}. Therefore, the RPC efficiency in the REFF simulation is calculated as the ratio between the number of avalanches with a size at the anode greater than 10$^{6}$ electrons and the total number of simulated events.

The REFF simulation has also been validated in 55\% C$_{3}$H$_{2}$F$_{4}$ and the remaining fraction of Ar, CO$_{2}$ or O$_{2}$. Since no measurements of $\alpha$\textsubscript{\textit{eff}} are available in such gas mixtures, the input values of $\alpha$\textsubscript{\textit{eff}} as a function of the \textit{HV} are obtained by a dedicated Monte Carlo simulation \cite{bianchi_PhDthesis} starting from the electron collision cross sections of C$_{3}$H$_{2}$F$_{4}$, Ar, CO$_{2}$ and O$_{2}$. Figure \ref{fig:plot0}b shows the comparison of the measurement with the simulation when the RPC efficiency is evaluated with $\eta$ = 0 mm$^{-1}$ and 10 mm$^{-1}$ in 55\% C$_{3}$H$_{2}$F$_{4}$ and 45\% Ar. A better agreement is obtained if $\eta$ is assumed to be 10 mm$^{-1}$, as shown in figure \ref{fig:plot0}a for the gas mixture 90\% C$_{2}$H$_{2}$F$_{4}$ and 10\% \textit{i}-C$_{4}$H$_{10}$. Hereafter, the value of $\eta$ is assumed to be 10 mm$^{-1}$ in the following simulations.

\section{Comparison of simulation and measurement of RPC efficiency in C$_{3}$H$_{2}$F$_{4}$-based gas mixtures} \label{sec:confrontodatiesimulazione}
Several studies suggest that C$_{3}$H$_{2}$F$_{4}$-based gas mixtures with the addition of CO$_{2}$ or He might be the most promising RPC gas mixtures with low GWP in terms of operating voltage, streamer probability and cluster size \cite{bianchi2019characterization_ok, Abbrescia_2016_He_ok}. Therefore, the agreement between the REFF simulation results and the experimental data are here evaluated for such type of gas mixtures; in particular, gas mixtures of C$_{3}$H$_{2}$F$_{4}$/CO$_{2}$ with and without \textit{i}-C$_{4}$H$_{10}$ and gas mixtures of C$_{3}$H$_{2}$F$_{4}$/He with a small fraction of \textit{i}-C$_{4}$H$_{10}$. In all cases values of $\alpha$\textsubscript{\textit{eff}}($HV$) are calculated by the same Monte Carlo simulation \cite{bianchi_PhDthesis} used for the validation of the REFF simulation in C$_{3}$H$_{2}$F$_{4}$-based gas mixtures.

\subsection{Gas mixtures of C$_{3}$H$_{2}$F$_{4}$/CO$_{2}$ with and without \textit{i}-C$_{4}$H$_{10}$}
\begin{figure}[h]
    \centering
    \includegraphics[width=0.65\textwidth]{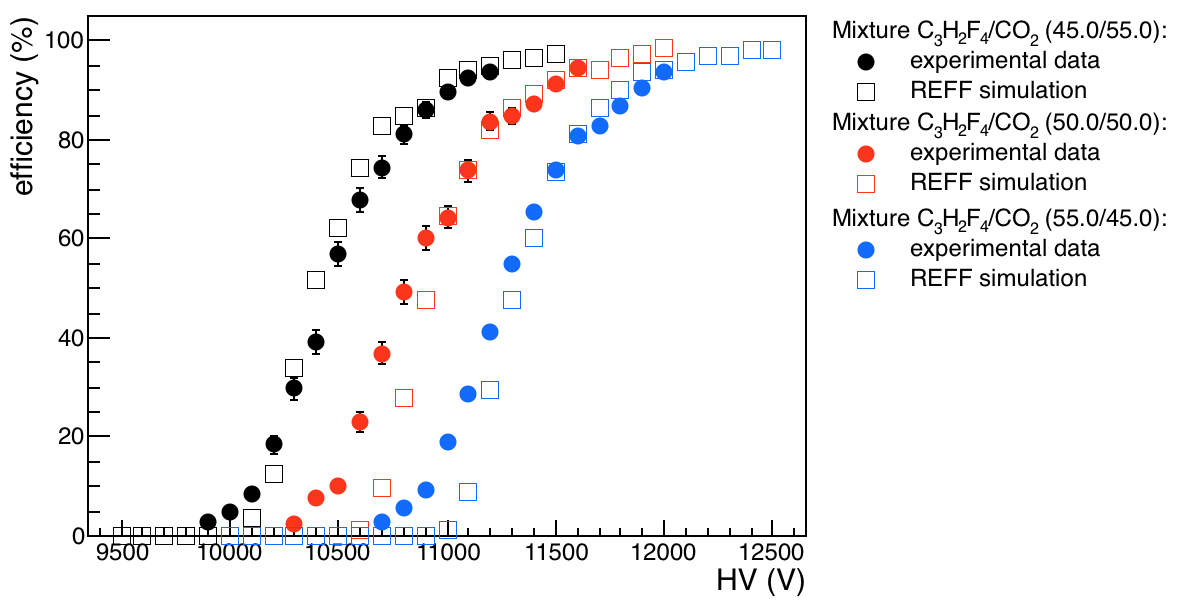}
    \caption{REFF simulation results and experimental data of RPC efficiency as a function of the $HV$ in C$_{3}$H$_{2}$F$_{4}$-based gas mixtures with the addition of CO$_{2}$ at different concentrations. Some statistical error bars are hidden by markers.}
    \label{fig:plot1}
\end{figure}

Figure \ref{fig:plot1} shows the simulation results and the experimental data of RPC efficiency as a function of the \textit{HV} with C$_{3}$H$_{2}$F$_{4}$ and CO$_{2}$ in different concentrations. The agreement between simulation and measurement of efficiency is good as well as the shift of the operating voltage, depending on the ratio of C$_{3}$H$_{2}$F$_{4}$/CO$_{2}$, is correctly described by the REFF simulation. For efficiency values greater than 70\%, the difference between simulation results and experimental data is lower than \textasciitilde10\%, whereas the difference of \textit{HV} values at efficiency of 50\% is lower than \textasciitilde100 V.

\begin{figure}[h]
    \centering
    \includegraphics[width=0.70\textwidth]{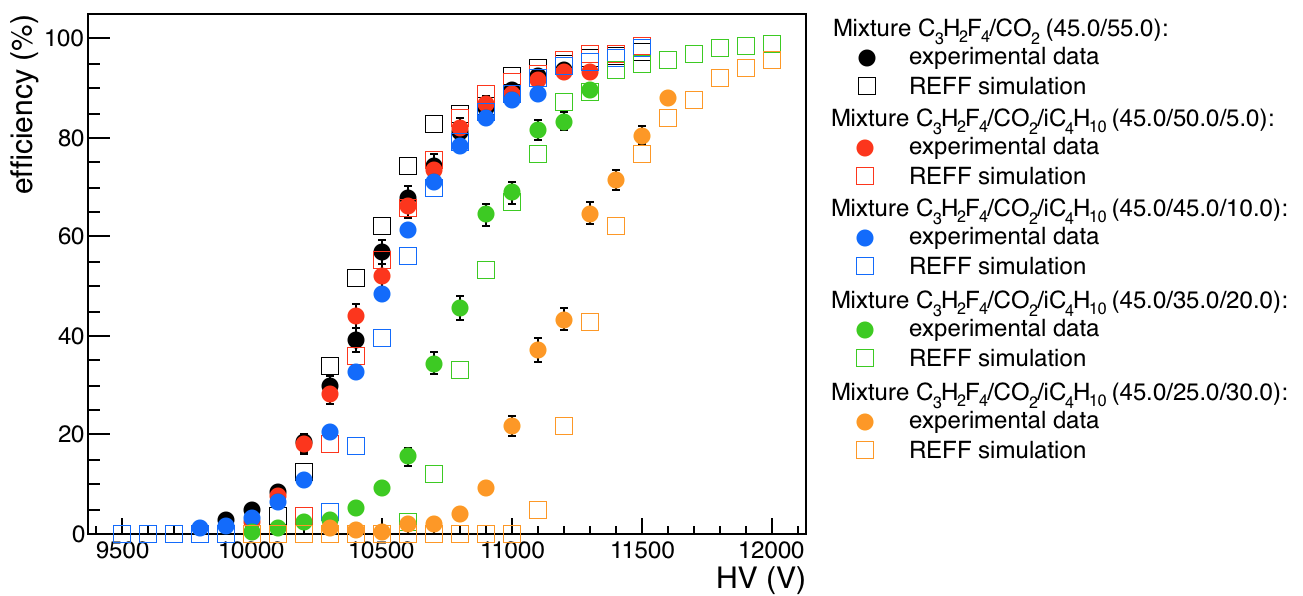}
    \caption{REFF simulation results and experimental data of detector efficiency as a function of the $HV$ in gas mixtures of 45\% C$_{3}$H$_{2}$F$_{4}$ and different concentrations of CO$_{2}$ and \textit{i}-C$_{4}$H$_{10}$. Some statistical error bars are hidden by markers.}
    \label{fig:plot2}
\end{figure}

Figure \ref{fig:plot2} shows efficiency curves as a function of the $HV$ with 45\% C$_{3}$H$_{2}$F$_{4}$ and different concentrations of CO$_{2}$ and \textit{i}-C$_{4}$H$_{10}$. In this case efficiency curves do not turn out to be shifted with \textit{i}-C$_{4}$H$_{10}$ fractions lower than 10\%. On the contrary, the shift of efficiency curves is observed if more than 10\% \textit{i}-C$_{4}$H$_{10}$ is added in place of CO$_{2}$. Similarly to the previous case, the shift of efficiency curves due to the variation of the ratio between CO$_{2}$ and \textit{i}-C$_{4}$H$_{10}$ is well reproduced by the REFF simulation. For efficiency values greater than 70\%, the difference between simulation and measurement is lower than \textasciitilde10\%, whereas the difference of $HV$ values at efficiency of 50\% is lower than \textasciitilde100 V. 

\subsection{Gas mixtures of C$_{3}$H$_{2}$F$_{4}$ and He}
\begin{figure}[h]
    \centering
    \includegraphics[width=0.70\textwidth]{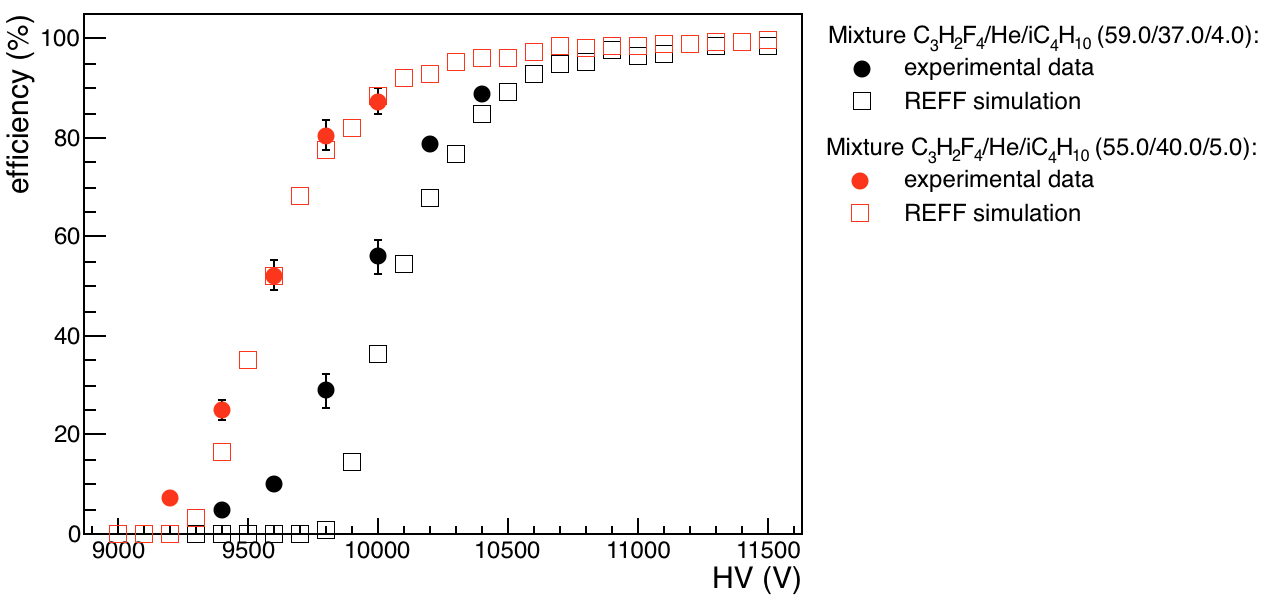}
    \caption{REFF simulation results and experimental data of RPC efficiency as a function of the $HV$ values in C$_{3}$H$_{2}$F$_{4}$-based gas mixtures with He and \textit{i}-C$_{4}$H$_{10}$. Measurements are provided by Abbrescia et al. \cite{Abbrescia_2016_He_ok}. Some statistical error bars are hidden by markers.}
    \label{fig:plot3}
\end{figure}

Results of the REFF simulation are also in agreement with measurements taken with different experimental set-up and front-end electronics. In particular, the comparison is here done in gas mixtures of C$_{3}$H$_{2}$F$_{4}$ with the addition of He and a small fraction of \textit{i}-C$_{4}$H$_{10}$. In this case measurements are taken by Abbrescia et al. in a 2 mm single-gap RPC with bakelite electrodes at 990 mbar and 293 K while the discrimination threshold for the data acquisition is \textasciitilde300 fC \cite{Abbrescia_2016_He_ok}. These parameters are appropriately taken into account for the simulation of the RPC efficiency. Figure \ref{fig:plot3} shows simulation results and experimental data of efficiency as a function of the $HV$ in C$_{3}$H$_{2}$F$_{4}$-based gas mixtures with He and \textit{i}-C$_{4}$H$_{10}$. The agreement between simulation and measurement is very good in the gas mixture 55\% C$_{3}$H$_{2}$F$_{4}$, 40\% He and 5\% \textit{i}-C$_{4}$H$_{10}$, in fact the difference is lower than \textasciitilde10\% for all $HV$ values. Concerning the gas mixture 59\% C$_{3}$H$_{2}$F$_{4}$, 37\% He and 4\% \textit{i}-C$_{4}$H$_{10}$, the agreement between simulations and measurements is quite satisfactory, even if the efficiency curve is not exactly reproduced by the REFF simulation. Nevertheless, the difference between simulation results and experimental data is lower than \textasciitilde15\% for efficiency values greater than 70\%, whereas the difference of $HV$ values at efficiency of 50\% is lower than \textasciitilde100 V.

\section{Conclusions}\label{sec:conclusioni}
We have presented a simulation procedure (REFF simulation) to calculate the efficiency of RPCs with several C$_{3}$H$_{2}$F$_{4}$-based gas mixtures and the addition of various gases. In this contribution comparisons of simulation and measurement are only reported for the most promising RPC gas mixtures based on C$_{3}$H$_{2}$F$_{4}$, however we obtained reliable simulations of the RPC efficiency in different combinations of C$_{3}$H$_{2}$F$_{4}$, Ar, CO$_{2}$, He, \textit{i}-C$_{4}$H$_{10}$ and O$_{2}$ in various concentrations \cite{bianchi_PhDthesis}. This highlights just how versatile the REFF simulation is for the identification of the most promising C$_{3}$H$_{2}$F$_{4}$-based gas mixtures with a low GWP for RPCs. However, some limitations of this simulation have been found in presence of SF$_{6}$ in the gas mixture. Research into solving this issue is already underway \cite{bianchi_PhDthesis}. In parallel, we are currently investigating possible future improvements of our simulation in order to estimate the RPC performance with C$_{3}$H$_{2}$F$_{4}$-based gas mixtures in terms of streamer probability, cluster size and time resolution.


%\newpage

\end{document}